\documentclass[aps,pra,showpacs,twocolumn]{revtex4}
\usepackage{graphicx}
\usepackage{bm}% bold math
\usepackage{amsmath}% equation formating
\usepackage{esvect}

\providecommand{\abs}[1]{\left\lvert#1\right\rvert}

\providecommand{\bra}[1]{\langle #1 \rvert}
\providecommand{\ket}[1]{\lvert #1 \rangle}

\providecommand{\be}{\begin{equation}}
\providecommand{\ee}{\end{equation}}
\providecommand{\ba}{\begin{eqnarray}}
\providecommand{\ea}{\end{eqnarray}}

\usepackage{mathbbol}

\usepackage{tikz}
\usepackage{amsfonts,amssymb}
\usepackage{dsfont}
\usepackage{physics}
\usepackage{tocvsec2}

\usepackage{appendix}

\begin{document}

\title{Time-frequency as quantum continuous variables}
 
\author{Nicolas Fabre\footnote{nfabre@ucm.es}$^{1}$, Camille No\^us$^2$, Arne Keller$^{3,4}$ and P\'erola Milman$^{3}$}

\affiliation{$^{1}$Departamento de Óptica, Facultad de Física, Universidad Complutense, 28040 Madrid, Spain}
\affiliation{$^{2}$ Laboratoire Cogitamus}
\affiliation{$^{3}$Université de Paris, CNRS, Laboratoire Matériaux et Phénomènes Quantiques, 75013 Paris, France}
\affiliation{$^{4}$ Department de Physique, Université Paris-Saclay, 91405 Orsay Cedex, France}

\begin{abstract}
We present a second quantization description of frequency-based continuous variables quantum computation in the subspace of single photons. For this, we define frequency and time operators using the free field Hamiltonian and its Fourier transform, and show that these observables, when restricted to the one photon per mode subspace, reproduce the canonical position-momentum commutation relations. As a consequence, frequency and time operators can be used to define a universal set of gates in this particular subspace. We discuss the physical implementation of these gates as well as their effect on single photon states, and show that frequency and time variables can also be used to implement continuous variables quantum information protocols, in the same way than polarization is currently used as a two-dimensional quantum variable. 
\end{abstract}
\vskip2pc 
 
\maketitle

\section{Introduction}

Since its beginnings and more than ever nowadays, the perspectives opened by using quantum mechanics in processing information and securing communication have excited the scientific community. The quantum computer - whatever meaning one gives to it -, that about fifteen years ago appeared as a distant mirage and an almost imaginary device, seems closer and closer to reality, materializing and capitalizing in all the senses of the word, in a scalable way, the superior performance of some quantum algorithms. Apart from the advertised scientific and non-scientific implications of quantum information research, it seems reasonable to think that information processing machines must make a different use of quantum physics  - and consequently, technologies - than the old-fashioned computer we are using to type these words. Also, in spite of the relevance, one thinks such technological breakthroughs have to fundamental research, we can safely state that quantum information changed the way one approaches not only information theory, but also quantum physics in general and, in particular, quantum optics.

Quantum field statistics, the field intensity, or its modal decomposition - in summary, all possible ways of measuring, manipulating and extracting information from radiation - are the basic tools of different quantum computation, and communication architectures based in quantum optics. Starting from a generic perspective, information can be extracted from, carried or processed by quantum fields by describing them essentially in two distinct ways: using observables with a continuous or with a discrete spectrum. Of course, both regimes can also be combined into hybrid devices \cite{Hybrid}, which gather the advantages and drawbacks of continuous and discrete variables.

In quantum optics, the traditional examples of continuous variables are the field's quadratures - linear combinations of the electric and magnetic field's amplitudes -, whose measurement using, for instance, homodyne detection, can provide complete information about the quantum state of radiation, encoded in the field's statistics. Alternatively, one can perform measurements that count \cite{sridhar_direct_2014,zhong_phase-programmable_2021,hamilton_gaussian_2017}  and correlate photons. The photon number is often associated to encoding information in discrete variables, represented by the photon number itself or the presence or absence of photons. It is also possible to use single photons degrees of freedom - as the polarization or the orbital angular momentum - as  discrete variables, and this is particularly useful in quantum communication and in testing fundamental aspects of quantum mechanics \cite{aspect_experimental_1982,PhysRevLett.115.250401,PhysRevLett.115.250402}.

Single photon's continuous degrees of freedom, as their longitudinal momentum or transverse position and momentum can be associated to continuous variables, also leading to an alternative implementation of quantum computation \cite{tasca_continuous_2011}, quantum communication \cite{walborn_quantum_2007} and fundamental tests of quantum mechanics \cite{BellWithMom}. In \cite{tasca_continuous_2011}, in particular, the isomorphism between the propagation equation of single photons  transverse variables and the Schr\"odinger equation of a massive particle was used to define continuous variables universal logical gates manipulating the transverse momentum and position of the photon's wave function with simple linear quantum optical devices or spatial light modulators (SLM).

In the present paper, we provide a comprehensive second quantization based description of frequency and time variables of single photons, showing how to manipulate and use these degrees of freedom as quantum continuous variables. For such, we introduce universal quantum  operators acting on a subspace consisting of single photons which occupy distinct modes. These gates are perfectly analogous to the ones introduced in \cite{lloyd_quantum_1999,braunstein_quantum_2005} for generic position and momentum-like variables. Frequency and time variables do not obey, in the general case, the same commutation relation as position and momentum. Nevertheless, in the subspace where each mode is at most occupied by one photon, such commutation relations can be retrieved from the bosonic commutation relations and provided that we consider frequency modes in experimentally reasonable ranges, avoiding for instance the issue of negative frequencies. 

Once these commutation relations between operators associated to time and frequency measurements of single photons are obtained, they can be used to define a set of universal gates in the same form as the ones manipulating momentum and position variables or the field's quadratures. This is the main scope of the present paper. We'll then introduce the physical contexts where such a universal set can be implemented, as well as its properties and ways to represent and detect quantum information encoded in frequency and time. In particular, we'll see that even though highly non-linear frequency and time gates are required, it is possible to define a quantum information processing architecture both in a measurement-based and in the circuit model with current technology. Finally, we'll discuss the phase space representation of frequency and time as quantum continuous variables, which is a useful tool to provide intuition on some non-classical aspects of states. We'll see however that the non-classical picture given by the constructed phase space is different from the one of quadratures.  \\

The paper is organized as follows. In Sec.~\ref{Intro}, we define the spectral and temporal properties of single photons through a Hilbert space description. We underline the possibility of defining time and frequency operators and the importance of being in the single-photon subspace. In Sec.~\ref{SecUniversal}, the universal set of gates for time and frequency continuous variables is defined based on the mathematical analogy with quadrature position-momentum variables. Experimental implementations are also discussed. In Sec.~\ref{PhaseSpace}, we recall the principles of the chronocyclic phase space description of single photons and consider in details the physical interpretation of such a phase space.

\section{States, operators and commutation relations} \label{Intro}

\subsection{Preliminary tools}

A single photon pure state at mode $i$ with frequency $\omega$ is described by the application of the creation operator to the vacuum state: $\hat{a}_{i}^{\dagger}(\omega)\ket{\text{vac}}=\ket{\omega}_i$. The label $i$ can be polarization, a spatial mode - as the transverse propagation direction -, or any other combination of modes that plays the role of an ancillary mode that creates distinguishability between each photon.  We can of course also define the annihilation operator such that $\hat{a}_i(\omega)\ket{\omega'}_i=\delta(\omega-\omega')\ket{\text{vac}}$. In addition, the commutation relation between creation and annihilation operator can be written as:
 
 \begin{equation}\label{commutationrl}
 [\hat{a}_{\alpha}(\omega),\hat{a}_{\beta}^{\dagger}(\omega')]=\delta(\omega-\omega')\delta_{\alpha\beta}\mathds{I},
 \end{equation}
 where $\alpha$ and $\beta$ are auxiliary modes. We also have that   $[\hat{a}_{\alpha}(\omega),\hat{a}_{\beta}(\omega')]=0$ and $[\hat{a}^{\dagger}_{\alpha}(\omega),\hat{a}^{\dagger}_{\beta}(\omega')]=0$.

If we consider to be in the narrow-band approximation \cite{PhysRevA.72.032110}, so that the central frequency of the spectral distribution is much larger than its spectral width, integrals can be extended over the whole frequency spectrum, and the Fourier transform of the annihilation operator is the annihilation operator at the arrival time $t$:

 \begin{equation}\label{Fourier}
 \hat{a}(t)=\frac{1}{\sqrt{2 \pi}}\int_{\mathds{R}} d\omega \hat{a}(\omega) e^{-i\omega t},
 \end{equation}
 the same being valid for the creation operation, of course. We also have that 
  \begin{equation}\label{commutationrltime}
 [\hat{a}_{\alpha}(t),\hat{a}_{\beta}^{\dagger}(t')]=\delta(t-t')\delta_{\alpha\beta}\mathds{I},
 \end{equation}
 where $\alpha$ and $\beta$ are auxiliary modes, and   $[\hat{a}_{\alpha}(t),\hat{a}_{\beta}(t')]=0$ and $[\hat{a}^{\dagger}_{\alpha}(t),\hat{a}^{\dagger}_{\beta}(t')]=0$.

 We stress that in the present description time is seen not as a parameter but as a degree of freedom associated to the  arrival time of photons in a detector.

 \subsection{States}
 
 A general single photon pure state can be decomposed in the time  basis or, equivalently, in the spectral basis as,
\begin{equation}\label{state}
\ket{\psi}=\int_{\mathds{R}} d\omega S(\omega) \hat{a}^{\dagger}(\omega)\ket{\text{vac}}.
\end{equation}
The spectrum $S(\omega)$ is the Fourier transform of the time of arrival distribution and $\abs{S(\omega)}^{2}=\abs{\bra{\omega}\ket{\psi}}^{2}$ denotes the probability density of detecting a photon with frequency $\omega$.  We can, of course, also construct from this principles  general mixed single photon states described by a density matrix.

The space of states we consider in the present contribution consists of a collection of $n$ single photon states in $n$ different ancillary modes. This space will be called from now on ${\cal S}_n$, where $n$ is the number of distinguishable modes and also the number of photons. It means that only cases where there is at most one photon per mode are considered. 

A general pure state in ${\cal S}_n$  can be written as
\begin{equation}\label{State}
\ket{\psi}=\int d\omega_{1}...\int d\omega_{n}  F(\omega_{1},...,\omega_{n}) \hat{a}_{1}^{\dagger}(\omega_{1})...\hat{a}_{n}^{\dagger}(\omega_{n})\ket{0},
\end{equation}
where the spectral function $F(\omega_{1},...,\omega_{n})$ is normalized to one : $\int d\omega_{1}...\int d\omega_{n} | F(\omega_{1},...,\omega_{n})|^2 =1$. One of the goals of defining a set of universal gates is to show that states as (\ref{State}) can be approached with arbitrary precision, from an arbitrary initial state using a finite set of operations. 

Of course, for $n \geq 2$, we'll have to consider, in addition to separable states, frequency and time entangled states.  In order to introduce the basics of entanglement in ${\cal S}_n$, let's discuss one of the simplest and most currently processes used to produce pairs of entangled  states, which is  spontaneous parametric down conversion (SPDC). In this case, we have that the two-photon output state can be written as :  

\begin{equation}\label{SPDC}
\ket{\psi}=\iint d\omega_{s}d\omega_{i} \text{JSA}(\omega_{s},\omega_{i}) \hat{a}_s^{\dagger}(\omega_{s})\hat{a}_i^{\dagger}(\omega_{i}) \ket{\text{vac}},
\end{equation}
where we considered that the creation operators of signal and idler photons are distinguishable and associated to different auxiliary modes labeled $s$ and $i$. The JSA acronym denotes ``joint spectral amplitude". One can also define its Fourier transform, the joint temporal amplitude, or JTA. The JSA contains all the information about the two photon frequency state.

Of particular interest for analyzing entanglement in state Eq.~(\ref{SPDC}) is its decomposition in a set of orthogonal functions, as introduced in \cite{PhysRevLett.84.5304,walborn_generalized_2012,lamata_dealing_2005}. One can show that
\begin{equation}\label{Schmidt}
\ket{\psi} = \sum_{n=1}^{\infty} \Lambda_{n} \ket{f_{n}}_{a}\otimes\ket{g_{n}}_{b}.
\end{equation}
An effective Schmidt rank $K_N$ can be defined as $1/K_N = \sum_{n=1}^{N}\Lambda_n^2$. If there is an $N$ such that
$K_N>1$, we can conclude that the state $\ket{\psi}$ is entangled.

We can study in details the particular case of a Gaussian shaped JSA, corresponding to the Joint Spectral Intensity (JSI) shown in Fig.~\ref{illustrationbloch}.The JSI has no phase information, but displays the essential information for analyzing correlations - and entanglement, when provided with phase information - between two photons. In the context of SPDC, the JSI's shape, and the existence of different degrees of frequency correlation between signal and idler photons, depend on the energy and momentum conservation of the non-linear interaction generating the photon pair, in one hand, and the pump laser spectral properties on the other hand. More specifically, the energy conservation condition is related to the sum of frequencies of signal and idler photons while the phase-matching condition is related to the difference of frequencies \cite{Orieux:11}.

Fig.~\ref{illustrationbloch} (a) displays a situation where the spectrum of signal and idler photons are uncorrelated, so the Schmidt decomposition has only one term, $\Lambda_1 \ket{f_1}_a\ket{g_1}_b$. It corresponds to a separable state of two photons, and this can be seen from the circular symmetry of the JSI, which means that it can always be decomposed as a product of two functions of any pair of variables related to $\omega_s$ and $\omega_i$ by an unitary transformation: the circle has no privileged principal axis. 

Notice that this situation is different from the one represented in Fig.~\ref{illustrationbloch} (b), where the JSI displays anti-correlation between variables $\omega_s$ and $\omega_i$.  As a matter of fact, the JSI in Fig.~\ref{illustrationbloch} (b), is an ellipsoid with principal axis $\omega_{\pm}=(\omega_s \pm \omega_i)/\sqrt{2}$, which means that the JSI can be written as product of two independent functions of these two variables. By defining $\delta_{\pm}$ as the width of the distribution in the $\omega_{\pm} = (\omega_s \pm \omega_i)/\sqrt{2}$ axis, we have that the ratio $r=\delta_+/\delta_- < 1$. It's a known result that the degree of squeezing of the JSA, and consequently of the JSI, is proportional to the degree of mode entanglement for pure quantum states \cite{lamata_dealing_2005}: the Schmidt rank $K_N$ (or alternatively, the degree of mode entanglement with the chosen mode partition) increases with decreasing $r$.

From the discussion above, we see that the JSA associated to the JSI in Fig.~\ref{illustrationbloch} (b) cannot be separated in a product of functions of variables $\omega_s$ and $\omega_i$ only, and these variables are correlated.  Nevertheless, a change of variables in the JSA to $\omega_{\pm}$ instead of $\omega_s$ and $\omega_i$ transforms it into a separable function. If one could, in addition to that, associate signal and idler photons as well to frequencies $\omega_+$ and $\omega_- $ instead of $\omega_s$ and $\omega_i$, then the Schmidt decomposition would have a single term and would correspond to a product state. This is an important point for the proposed quantum computation set-up that will be discussed in details in Sec.~\ref{SecUniversal}. Before this, we introduce time and frequency operators which will be shown to be analogous to the quadratures of the electromagnetic field, expressed here in the simplified dimensionless form $\hat p=i(\hat a^{\dagger}-\hat a)/2$ and $\hat x=(\hat a^{\dagger}+\hat a)/2$. In this sense, frequency and time can also be considered as continuous variables to encode quantum information.

\begin{figure}\label{fig1}
\end{figure} 

\subsection{Time and frequency operators}

The time and frequency operators are defined as:

\begin{align}
\hat{t}_{a}=\int_{\mathds{R}} t \hat{a}^{\dagger}(t)\hat{a}(t) dt,\\
\hat{\omega}_{a}=\int_{\mathds{R}} \omega \hat{a}^{\dagger}(\omega)\hat{a}(\omega) d\omega.
\end{align}
When applied to single photons states, these operators fulfill the eigenvalues equation : $\hat{t}_{a}\ket{t}_{a}=t\ket{t}_{a}$ and $\hat{\omega}_{a}\ket{\omega}_{a}=\omega \ket{\omega}_{a}$. The frequency operator is proportional to the  free Hamiltonian $\hat{E}_{a}=\hbar \hat{\omega}_{a}$. 
As previously, we considered the narrow band approximation of a photon with central frequency far from origin. Consequently, the integration over the frequency can safely be considered as covering all $\mathds{R}$. 
As for the time variable, it corresponds to the Fourier transform of frequency for all practical purposes and is physically associated to the time of detection conditioned to the fact that a detection has indeed happened \cite{giovannetti_quantum_2015,maccone_quantum_2020}. 

Using Eq.~(\ref{commutationrl}), we can see that time and frequency operators do not commute in the single photon  (single mode) regime (see Appendix \ref{Expansion}):

 \begin{equation}\label{noncommutation}
 [\hat{\omega}_{a},\hat{t}_{a}]=i\mathds{I}. 
 \end{equation}
They form, together with the identity operator $\mathds{I}$, a three-dimensional Heisenberg algebra in perfect analogy with the position and momentum operators. This fact, not true in general for modes occupied by more than one photon (see Appendix A), is essential for building a set of universal gates which manipulate frequency and time as the universal gates defined for position and momentum manipulate states defined in these basis. The present manuscript is thus entirely based on considering $\hat x \rightarrow \hat \omega$ and $\hat p \rightarrow \hat t$.

Also, from Eq.~(\ref{noncommutation}), we can use the Robertson's inequality to derive a time-energy Heisenberg inequality:

\begin{equation}\label{Robertson}
\Delta \hat{E}_{a}\Delta \hat{t}_{a} \geq \hbar/2.
\end{equation}
Rather than expressing a classical Fourier transform relation, Eq.~(\ref{Robertson}) provides an operator description of the time and the frequency bandwidth at the single photon level as quantum noise \cite{fadel_time-energy_2021}.

\section{Universal operations}\label{SecUniversal}

We have introduced in the previous section the basic tools which will be useful for the definition of a universal set of quantum operations using time and frequency as continuous variables. 

Before moving to its description, we recall the basics of continuous variables universality, as introduced in \cite{braunstein_quantum_2005}. The authors show that it's possible to define an universal set of unitary operators using observables $\hat x$ and $\hat p$, and central to their result is the fact that such observables obey the commutation relation $\left [\hat x, \hat p \right ]= i\mathds{I}$ (where we have used dimensionless operators for convenience). Such operators can be associated either to the position and momentum variables of a particle or, in the context of quantum optics, to the electromagnetic field's quadrature. 

A universal set of unitary operators is defined as a finite set that, when combined, approaches with arbitrary precision any unitary operation which is a polynomial in $\hat x$ and $\hat p$. There are of course many possible sets, but we'll focus on the following one : 
\begin{equation}\label{Universal}
{\cal U} = \{e^{i\hat x s}, e^{i\frac{\pi}{4}(\hat x^2 + \hat p^2)},e^{i \hat x^2 s}, e^{i \hat x_i \otimes \hat x_j}, e^{i\hat x^3 s} \}.
\end{equation}
We briefly explain the action and properties of each one of the unitary operators in ${\cal U}$ :  $e^{i\hat x s}$ is the displacement operator that displaces momentum of an amount of $s$. $e^{i\frac{\pi}{4}(\hat x^2 + \hat p^2)}$ is the Fourier transform, and $e^{i \hat x^2 s}$ is called a shear operation (governed by the parameter $s$), that compresses one quadrature while also implementing a quadrature rotation. As for $e^{i \hat x_i \otimes \hat x_j}$, it involves two distinct modes, $i$ and $j$, and is an entangling operation that generates displacements in one mode according to the state of the system in the other mode. Finally, $e^{i\hat x^3 s}$ is a non-Gaussian operation that is essential to generate polynomials of order higher than $2$ using the canonical commutation relation. 

A first comment is that the set ${\cal U}$ acts on what one calls qumodes, which are, in the case of the field's quadratures, frequency, momentum or polarization distinct modes. Each operation is thus defined independently for each mode, except for $e^{i \hat x_i \otimes \hat x_j}$ that couples two different modes. 

As mentioned, universality of the set ${\cal U}$ is proven using the fact that $\hat x$ and $\hat p$ obey the canonical commutation relation. Any other pairs of observables obeying the same commutation relation can be used to define a universal set of operations in the same form as the ones in ${\cal U}$. Consequently, using such operators, it's possible to generate a complete computational space, provided that one identifies a  set-up where such operators make a physical sense and can be implemented. 

From these observations, we move to the time and frequency variables restricted to the single photons subspace. Using  Eq.~(\ref{noncommutation}), we deduce that with the set : 
\begin{equation}\label{UniversalFreq}
{\cal W} = \{e^{i\hat \omega s}, e^{i\frac{\pi}{4}(\hat \omega^2 + \hat t^2)},e^{i \hat \omega^2 s}, e^{i \hat \omega_i \otimes \hat \omega_j}, e^{i\hat \omega^3 s} \}
\end{equation}
we are capable of approaching, with arbitrary precision, any operation within the $n$ single photons subspace ${\cal S}_n$ (in analogy with the $n$ qumodes one), {\it i.e.}, we are capable of constructing unitary operators which approach with arbitrary precision any polynomial in $\hat \omega$ and $\hat t$ in ${\cal S}_n$. 

The set ${\cal W}$ is then a universal one, in the sense of \cite{braunstein_quantum_2005}, in ${\cal S}_n$, for time and frequency operations.   

A first comment is that all the operators in ${\cal W}$ are non-linear in $\hat a(\omega), \hat a^{\dagger}(\omega)$, the annihilation and creation operator in a given mode $\omega$. $\hat a^{\dagger}(\omega)\hat a(\omega)$ is the field's intensity in this mode, which is itself a polynomial in $\hat x(\omega) = \frac{\hat a(\omega)^{\dagger}+\hat a(\omega)}{\sqrt{2}}$ and $\hat p(\omega)=i \frac{\hat a(\omega)^{\dagger}-\hat a(\omega)}{\sqrt{2}}$ and, for this reason, can be built from ${\cal U}$, as is the case for all operators in ${\cal W}$. 

With this in mind, that defines the computational space, we now briefly discuss the operators  in ${\cal W}$, their physical meaning and different ways to implement them.  

\subsection{The displacement operator and the Fourier transform}

The operator $e^{i\hat \omega_{a} s}$ is the generator of displacement in time of a single photon in mode $a$, and it can be associated to the free evolution. Restricted to the ${\cal S}_n$ subspace, the expansion of the exponential leads to (see Appendix \ref{Expansion} and \cite{PhysRevA.102.012607})
\begin{equation}
{\cal{\hat{D}}}_{\hat{\omega}_{a}}(s)=e^{i\hat \omega_{a} s}=\int \hat{a}^{\dagger}(t+s)\hat{a}(t) dt,
\end{equation}
which corresponds to the annihilation of one photon at a given time and the creation of a photon at a displaced time, as one would expect. Operator $\hat F = e^{i\frac{\pi}{4}(\hat \omega_{a}^2 + \hat t_{a}^2)}$ acts as a Fourier transform so that 
\begin{equation}
{\cal{\hat{D}}}_{\hat{t}_{a}}(\mu)=\hat F {\cal{\hat{D}}}_{\hat{\omega}_{a}}(s) \hat F^{\dagger} =  \int \hat{a}^{\dagger}(\omega+\mu)\hat{a}(\omega) d\omega.
\end{equation}
Operators ${\cal{\hat{D}}}_{\hat{t}_a}(\mu)$ and ${\cal{\hat{D}}}_{\hat{\omega}_{a}}(s)$  verify the Weyl's algebra,
\begin{equation}\label{Weylalgebra}
{\cal{\hat{D}}}_{\hat{\omega}_{a}}(s)\hat{D}_{\hat{t}_{a}}(\mu)=e^{is\mu/2} \hat{D}_{\hat{t}_{a}}(\mu){\cal{\hat{D}}}_{\hat{\omega}_{a}}(s),
\end{equation}
where we have again used the bosonic non-commutation relation Eq.~(\ref{commutationrl}) \cite{Note}.

We now discuss the optical implementation of the introduced operators. Time displacements in the single photon regime can be realized using the free evolution operator. As a matter of fact, since time is defined in practice by conditioning it to detection \cite{maccone_quantum_2020}, we are most of the time interested in time differences, which are associated to optical path differences that reveal the photonic temporal profile, as for instance in interferometers. As for frequency displacements, they require non-linear optics, as in  \cite{hu_-chip_2021}. This is an important difference from the implementation point of view with the previously mentioned quadrature-based continuous variable quantum computation scheme, where all displacement operators require using  linear optics only.

The Fourier transform $\hat F$ has a quadratic form in frequency and time. Such an operator, in the quadrature or transverse position/momentum based encoding, is proportional to the system's free Hamiltonian: the electromagnetic field's energy, in the quadrature case, or the the free propagation in the transverse position/momentum one. Thus, in this last set-up, the free propagation can be used to implement either the fractional or the complete Fourier transform, according to the propagation distance  \cite{walborn_spatial_2010}. In the present frequency/time encoding, as discussed, the free Hamiltonian creates time displacements and is linear in $\hat \omega$.  Nevertheless, propagation inside a linear media leads to group velocity dispersion, which is a currently used technique to implement the Fourier transform from frequency to time variables and vice-versa. The fractional Fourier transform can  be implemented using the same technique by controlling the propagation distance. Finally, another possible way to implement  both fractional and the full  Fourier transform is using a $4f$-like configuration for frequency and time, as detailed in  \cite{fabre:tel-03191301}, or by using an optical quantum memory \cite{mazelanik_temporal_2020}. 

\subsection{Squeezing-like and non-Gaussian operations}

The remaining single photon operators in ${\cal W}$ consist of frequency and time dependent quadratic and cubic phases applied to the single photon states. They can be re-expressed as follows (see, for technical details, Appendix \ref{Expansion}): 
\begin{eqnarray}\label{cube}
&& e^{i s \hat \omega_{a}^2}= \int e^{i s \omega^2}\hat a^{\dagger}(\omega)\hat a(\omega) {\rm d}\omega \nonumber, \\
&& e^{i \gamma \hat \omega_{a}^3}= \int e^{i \gamma \omega^3}\hat a^{\dagger}(\omega)\hat a(\omega) {\rm d}\omega.
\end{eqnarray}
 
These transformations, either in the frequency or in the time basis, can be realized using spatial light modulators after mapping frequency to transverse position \cite{mazzotta_high-order_2016} or, in a measurement-based model, by pump engineering \cite{boucher_toolbox_2015}. As a consequence, it's possible to implement them without using non-linearity at the single photon level. We'll leave the discussion about the meaning of squeezing and non-Gaussianity in the present encoding to Sec.~\ref{PhaseSpace}. 

\subsection{Two-photon gate}

We now move to  operator $ e^{i \hat \omega_i \otimes \hat \omega_j}$, which manipulates in a conditional way the frequency of two photons in distinct auxiliary modes. In order to build a more intuitive picture, we'll rather discuss the Fourier transform in mode $j$ of this operator, {\it i.e.}, $e^{i \hat \omega_i \otimes \hat t_j} = \hat F_j e^{i \hat \omega_i \otimes \hat \omega_j} \hat F_j^{\dagger}$. This operator acts in the state $\ket{\omega,\omega'}_{ij}$ as 
\begin{equation}\label{cnot}
e^{i \hat \omega_i \otimes \hat t_j} \ket{\omega,\omega'}_{ij} = e^{i  \omega  \hat t_j} \ket{\omega,\omega'}_{ij} =  \ket{\omega,\omega'-\omega}_{ij}.
\end{equation}
It is, indeed, a conditional operation that displaces the frequency of one photon of an amount that depends on the frequency of the other one, or a continuous variable version of the discrete CNOT gate. 

By combining this operator and other ones from the Gaussian set in ${\cal W}$ - as, again, the Fourier transform and the quadratic gate - one can build
\begin{eqnarray}\label{BS}
&&e^{i \frac{\pi}{4}(\hat \omega_i \otimes \hat t_j- \hat t_i \otimes \hat \omega_j}) = \\
&&\iint \hat a^{\dagger}(\frac{\omega_i+\omega_j}{\sqrt{2}})\hat a^{\dagger}(\frac{\omega_i-\omega_j}{\sqrt{2}})\hat a(\omega_i)\hat a(\omega_j){\rm d}\omega_i {\rm d}\omega_j. \nonumber
\end{eqnarray}
The applications of this operator are numerous \cite{PhysRevA.102.012607,fabre_spectral_2021} and present interesting analogies with the beam-splitter operator defined for continuous variables representation using quadratures. In the quadrature case, a balanced beam-splitter combines two modes of the electromagnetic field and performs a conditional operation that produces as an output, two modes which are the sum and the difference between two input modes. Still in the quadrature case, if we consider Gaussian states it's a known result that an optical circuit consisting of beam-splitters and passive linear optical elements manipulating modes \cite{fabre_modes_2020} can be used to physically implement the Bloch-Messiah decomposition \cite{braunstein_squeezing_2005} and transform mode entangled squeezed states into separable ones in the minimal mode representation of the considered state. In this scenario, beam-splitters play a central role, since they are two-mode operations that entangle and disentangle states by performing mode basis change.

 The transformation Eq.~(\ref{BS}) implements the same type of operation as beam-splitters perform to spatial modes and to the field's amplitude to the frequency degree of freedom. The frequencies of two input photons are transformed in a conditional way into the sum and difference of each photons' frequencies. It's natural to ask wether Eq.~(\ref{BS}) can play a similar role. For this, we refer again to the case of photon pairs generated by SPDC from a pump with a Gaussian spectrum, as in Sec.~\ref{Intro}. 

We have seen that in SPDC the frequency of the two generated photons can be correlated or anti-correlated according to the energy and momentum conservation functions of the pump \cite{francesconi_engineering_2020} and that the degree of mode entanglement of the photon pair depends on the parameter $r$. We'll show now that this spectral distribution is the perfect analog of a two-mode squeezed state in the quadrature basis and that it can be separated into a product state using a Bloch-Messiah like decomposition. 

This fact is not entirely obvious at first sight. It's true that for many relevant experimental set-ups, the interaction generating multi-mode entangled quadrature states in optical parametric oscillators (OPO) is the same as the one that generates photon pairs from SPDC. The main difference between these two physical situations is the intensity regime: while in OPOs multi-mode squeezed states are generated, in SPDC the interaction Hamiltonian is expanded until first order in the creation and annihilation operator, and only the photon pair generation process is considered. The produced state is then post-selected by detection, which ensures its non-Gaussian character and, with high probability, its single photon nature. Nevertheless, such intensity considerations do not change the properties of the JSA, and its expansion in a set of pairs of modes, as was done Eq.~(\ref{Schmidt}) and \cite{PhysRevLett.84.5304}, is exactly the same in both intensity regimes. In the case of OPOs, it leads to the mode decomposition that describes the state as a separable one (Bloch-Messiah decomposition) while in the case of SPDC it leads to the Schmidt decomposition. These two states, which are the result of the same mathematical operation in two different intensity regimes are, of course, essentially different.

\begin{figure*}
\begin{center}
\includegraphics[width=0.9\textwidth]{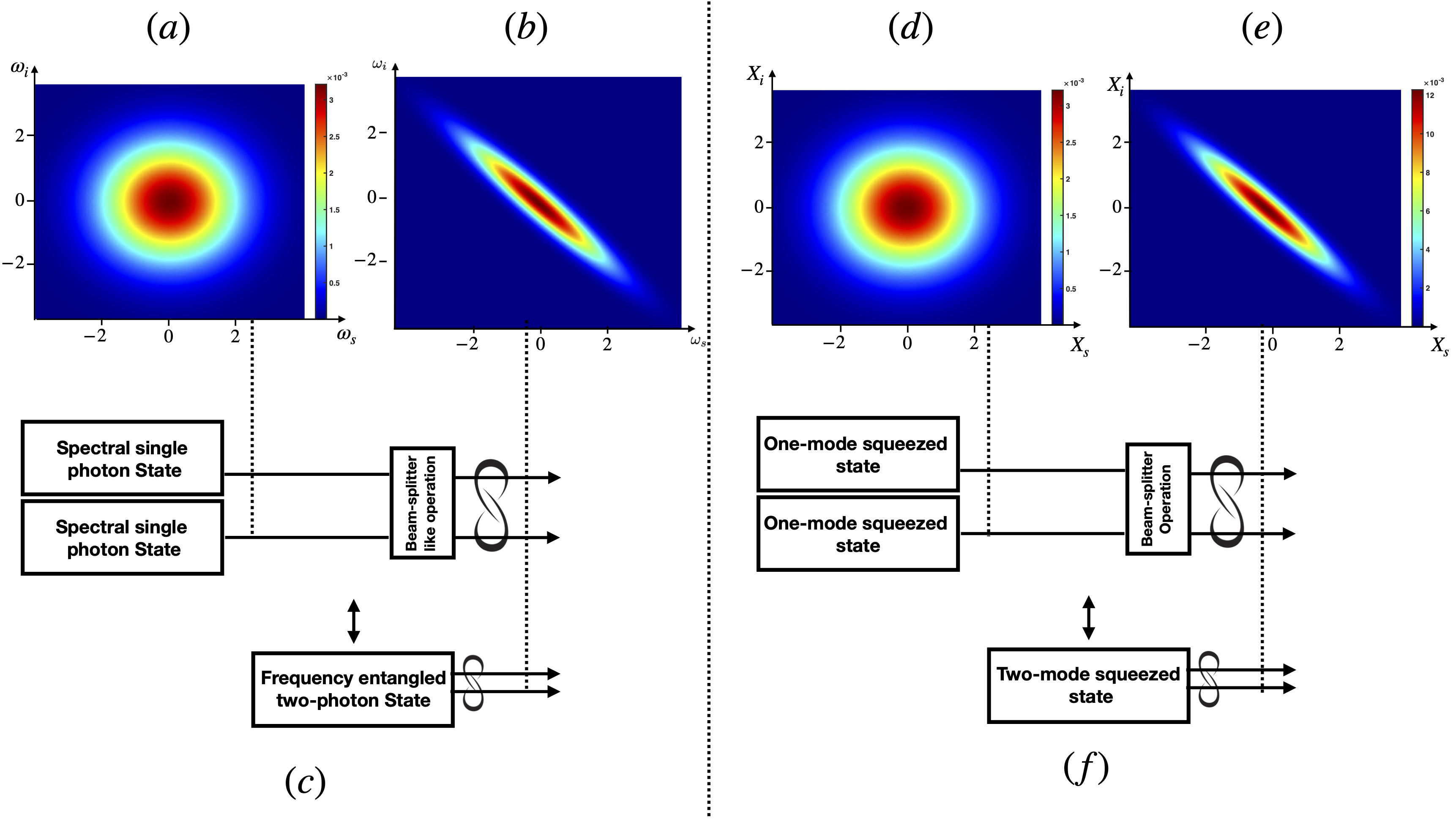}
\caption{\label{illustrationbloch}Illustration of Bloch-Messiah decomposition for  time-frequency (a) and quadrature (b) encoding. In the quadrature encoding, the beam-splitter interaction between two one-mode squeezed state leads to a two-mode squeezed state. In the time-frequency encoding, the analogous is the interaction of two single photons with a Gaussian spectrum through the beam-splitter -like operator Eq.~(\ref{BS}). The frequency unit have been rescaled with respect to a single photon state chosen as a reference. The ratio between the major and minor axis of the ellipse in (b) and (e) is here $1/10$. }
\end{center}
\end{figure*}

Nevertheless, they have the same JSA and if we consider it to be in the Gaussian form of Fig.~\ref{illustrationbloch} (b), it is a particular case of a function that can be expressed as a product of functions of only two different variables, $\omega_+$ and $\omega_-$. This fact can be expressed in the SPDC regime as:

\begin{align}\label{Separation}
\iint {\rm d}\omega_s {\rm d}\omega_i & \text{JSA}(\omega_s,\omega_i)\ket{\omega_s,\omega_i} = %\nonumber \\
 \iint {\rm d}\omega_+ {\rm d}\omega_-    f(\omega_+)g(\omega_-) \nonumber \\ \times  &\ket{(\omega_++\omega_-)/\sqrt{2},(\omega_+-\omega_-)/\sqrt{2}},
\end{align}

which is an entangled state. Now, it's clear that by applying Eq.~(\ref{BS}) to Eq.~(\ref{Separation}), we obtain: 

\begin{eqnarray}\label{Separation2}
&&e^{i \frac{\pi}{4}(\hat \omega_i \otimes \hat t_j- \hat t_i \otimes \hat \omega_j})\iint {\rm d}\omega_s {\rm d}\omega_s \text{JSA}(\omega_s,\omega_i)\ket{\omega_s,\omega_i}= \nonumber \\
&& \int {\rm d}\omega_+ f(\omega_+)\ket{\omega_+}\int {\rm d}\omega_- g(\omega_-)\ket{\omega_-}.
\end{eqnarray}
which is a separable state. 

So, we can conclude that in the single photon regime considered in the present contribution, the Bloch-Messiah decomposition also makes sense and transforms spectrally entangled Gaussian states into separable ones instead of transforming Gaussian entangled states (many-modes squeezed states) into separable ones, as is the case in the quadrature representation. Also, in the single photon regime considered here, the physical and mathematical implementation of the frequency beam-splitter like operator Eq.~(\ref{BS}) involve using a non-linear photon-photon interaction.This may sound unpractical, but this is not exactly the case: a particularly interesting consequence of this fact is that one can thus re-interpret the non-linear interaction that generates the photon pair as the application of gate Eq.~(\ref{BS}) to an initially separable state in variables  $\omega_s$ and $\omega_i$, suggesting that a measurement based (or hybrid) model of computation is probably the most suitable for the proposed configuration. This can be schematized as in the quantum circuit represented in Fig.~\ref{illustrationbloch}(c), and which is compared with the quadrature case Fig.~\ref{illustrationbloch}(d), (e), (f).  We should finally mention that another possible way to experimentally implement gate (19) gate is by using an auxiliary optical active medium such as atomic system \cite{PhysRevA.91.033816} or by using split-ring resonators \cite{kues_quantum_2019,yang_squeezed_2021,Lu:19,Grassani:15}.

As a conclusion, the Gaussian-shaped JSA can be factorized by the application of a beam-splitter like gate as Eq.~(\ref{BS}), in analogy to what happens in the Bloch-Messiah transformations for Gaussian states in quadratures.  Of course, the non-linear operations we defined modify, as in the quadrature case, the number of relevant modes used to express the state, since an entangled state involving many terms of the Schmidt decomposition is reduced to a single separable state. We recall that the nature of the physical implementation in the discussed example - a non-linear two-photon interaction - and the type of variables involved - the spectrum of single photons - are completely different from the quadrature case.

The presented scenario consists of an original way to analyse and interpret mode transformations that is not only relevant from the quantum information perspective but can also bring new applications of frequency-based quantum optics in the regime of individual photons. 

Finally, the case of a Gaussian-shaped function discussed here is by no means the only example of a state that can be separable in two variables, the presented techniques can be used for different distributions.

\subsection{Encoding a qubit with the continuous variables of single photons}

Universal quantum computing has to be supplemented with error-correction codes to provide fault-tolerant quantum computation. In the case of quadrature continuous variables encoding, different types of qubits have been defined, and as we see things today, the leading experimental and practical candidates are the so-called ``cat-code" \cite{cochrane_macroscopically_1999} and the Gottesman-Kitaev-Preskill (GKP) code \cite{gottesman_encoding_2001}. Each one of these codes is built so as to be resilient against different types of errors. Cat codes are robust against photon losses while GKP codes resist to small position and momentum displacements. A  detailed study of performance for these two codes can be found in \cite{albert_performance_2018}.

In the time-frequency encoding of single photons, similar codes can be defined. Time-frequency cat codes \cite{fabre_producing_2020}, defined as the linear superposition of two frequencies, are then built to be resilient against time displacement operations which is a second-order dispersive effect. Time-frequency GKP codes \cite{PhysRevA.102.012607} are protected against both small shifts in time and in frequency, and error correction procedures can be implemented using for instance the Steane protocol or teleportation based-error correction (see Ref.~\cite{fukui_all-optical_2021} in the quadrature case, which is the perfect analogous of the present one). Also, in certain optical fibers,  the frequency-polarization coupling \cite{gordon_pmd_2000,poon_polarization_2008,antonelli_pulse_2005} can induce frequency broadening. It's important to notice that the time-frequency encoding of single photons is conditioned to photon detection, so that photon losses do not affect the encoded information, but rather its existence contrary to the quadrature case.

\section{The chronocyclic phase space}\label{PhaseSpace}

In order to provide a full continuous variable description of frequency and time degrees of freedom of single photons, it's useful to introduce a phase space picture and interpret it. To begin with, it is indeed possible to define the chronocyclic phase space in the single photon subspace, as well as the chronocyclic Wigner distribution \cite{brecht_characterizing_2013,PhysRevA.102.012607}. As for the quadrature based phase space, this distribution has a one-to-one correspondance with the spectral density matrix. 

We consider continuous real variables, so the phase space is rectangular and we can apply the construction procedure relying on the Stratonovich-Weyl rules (see Ref.~\cite{tilma_wigner_2016} for a reminder of these rules). It starts from defining the parity operator

 \begin{equation}\label{parity}
\hat{\Pi}=\int \hat{a}^{\dagger}(\omega)\hat{a}(-\omega) d\omega
 \end{equation}
 with respect to the central frequency of a reference state that here was set to zero. The displaced parity operator can also be defined from Eq.~(\ref{parity}) as
 
\begin{equation}
\hat{\Pi}(\mu,\tau)=\hat{D}(\mu,\tau)\hat{\Pi}\hat{D}^{\dagger}(\mu,\tau),
\end{equation}
where $\hat{D}(\mu,\tau) = e^{-i\mu\tau/2} \hat{D}_t(\mu)\hat{D}_{\omega}(\tau)$ is a combined arbitrary time/frequency displacement. The expectation value of the displaced parity operator is the chronocyclic Wigner distribution:
\begin{equation}
W_{\hat{\rho}}(\mu, \tau)=\langle \hat{\Pi}(\mu,\tau) \rangle =  \int d\omega e^{2i\omega \tau} \bra{\mu-\omega}\hat{\rho}\ket{\mu+\omega}
\end{equation}
The chronocyclic Wigner function can be measured using a modal decomposition \cite{gil-lopez_universal_2021} or a frequency beam-splitter operation \cite{fabre_spectral_2021}, for instance. For a single photon pure state $\hat{\rho}=\ket{\psi}\bra{\psi}$ and $\ket{\psi}=\int S(\omega) d\omega\ket{\omega}$ , the chronocyclic Wigner distribution can be written as
 \begin{equation}\label{WignerPure}
 W_{\hat{\rho}}(\mu,\tau)=\int d\omega e^{2i\omega \tau}  S(\mu-\omega)S^{*}(\mu+\omega),
 \end{equation}
and this distribution can be  negative. In particular, it's always negative for pure non-Gaussian states. The marginals of the chronocyclic Wigner distribution correspond to the spectral and time-of arrival distributions, which can be measured directly experimentally :
\begin{align}
\int d\mu W_{\hat{\rho}}(\mu,\tau)= \abs{\tilde{S}(\tau)}^{2}, \nonumber \\
\int d\tau W_{\hat{\rho}}(\mu,\tau)= \abs{S(\mu)}^{2}. \nonumber
\end{align}

Considering displacements in the present case is essential, since there is no natural absolute reference state as the physical vacuum (the absence of photons): using a ``zero frequency" state as a reference doesn't make real sense since it's physically out of reach and doesn't correspond to energy scales that can be attained experimentally. We can thus re-scale the space with respect to a given reference state, with a given average energy and a finite Gaussian width in time and frequency. The origin of the phase space and the reference state is then  related to the physical situation of interest. Consequently, squeezing and relative displacements do not have a meaning {\it per se} and are defined with respect to a reference state.  Also, symmetries will be relative to some chosen axis in phase space, or equivalently, energy, that in all practical situations will be far from the zero frequency state. 
 
We can notice that the distribution Eq.~(\ref{WignerPure}) is quadratic in the spectral function of interest and its description is identical to the Wigner-Ville distribution for classical field \cite{dorrer_concepts_2005}. Based on these similarities, some may then conclude that the presented Wigner function does not display quantum properties of the photon. Nevertheless, this naïve inductive conclusion is wrong, and this can be shown using different types of arguments. The first one is also inductive, and compares frequency to polarization. Polarization is also a property that is well defined for classical fields (or any field statistics). Nevertheless, it is isomorphic to a two-dimensional  Hilbert space when associated to single photons, and this is a fundamental ingredient to demonstrate fundamental aspects of quantum mechanics using single photon polarization detection, as for instance in non-locality tests \cite{aspect_experimental_1982}. Frequency is the continuous analogous of polarization, and even if well defined for classical fields, it is a quantum (continuous) variable when associated to single photons. In spite of that, in the literature, frequency is usually discretized into modes or bins \cite{Qudits, PhysRevX.5.041017}. 
 
Given these facts, the difference between Gaussian and non-Gaussian distributions deserves some attention. Non-Gaussianity is known to be an essential property for continuous variables quantum computation, since it's essential to prove quantum advantage over classical protocols. Gaussian only quantum protocols can be efficiently simulated using classical resources. In the case of quadratures, it turns out that implementing the non-Gaussian operation in ${\cal U}$ is challenging. For instance, the quadrature cubic phase gate requires non-linearity and the production of ancillary non-Gaussian states which are hard to implement \cite{miyata_implementation_2016}.  However, this is not the case here: in the case of pure frequency-time states, the non-Gaussianity of the time-frequency  distributions is only related to some particular mode engineering, which leads to distributions with specific parity properties. For instance, all pure non-Gaussian states have negativities, or equivalently, their displaced parity is negative at some regions. This simply means that the spectrum of single photons in non-Gaussian states can be odd at some points, and such states are  a particular type of superposition states which are odd with respect to some displaced parity. One example is the state $\ket{\psi}=\frac{1}{\sqrt{2}}(\ket{\omega_1}-\ket{\omega_2})$) which is odd with respect to the axis $\omega_+=(\omega_1+\omega_2)/\sqrt{2}$. 

Of course, non-Gaussian states and operations in the present context are, as well, essential for universality, and the non-Gaussian operator in  ${\cal W}$ plays the same role as in  ${\cal U}$, which is transforming Gaussian operators into non-Gaussian ones. We recall that this is also the case when one uses polarization encoding, where  non-Clifford gates can be implemented using linear operations of the same nature of any other local gate in the universal set.  It's rather a particularity of quadrature-based continuous variables systems to have non-Gaussianity associated to a challenging experimental process. Instead of coming from a particular hard to implement physical operation, the fact that the present quantum computing architecture cannot be classically simulated comes from two independent facts. From the physics side, we have that we are dealing with single photons which obey to particular commutation relations. As a consequence, single photons prepared in different modes inherit quantum properties from these commutation relations, and this is the essential ingredient to manipulate, in a quantum way, every operator defined in ${\cal S}_n$. In particular, Gaussian and non-Gaussian operators. Thus, using single photon measurements and manipulations, ensures that we are dealing with a quantum system and that some of  the resulting operations cannot be efficiently simulated with classical ressources. Frequency, or polarization, or any other degree of freedom associated to single photons are used as ways to manipulate them, encoding and decoding quantum information and exploiting their fundamental symmetries.

Before concluding, we'll discuss the two-photon case so as to provide more intuition about entanglement and the phase space. The two-photon chronocyclic Wigner distribution can also be defined for pairs of single photon (${\cal S}_2$) and it provides information on both the amplitude and phase of the photon pair  \cite{brecht_characterizing_2013}:
 \begin{multline}
  W_{\hat{\rho}}(\omega_{s},\omega_{i},t_{s},t_{i})= \iint   e^{2i\omega'_{s}t_{s}}e^{2i\omega'_{i}t_{i}} d\omega'_{s}d\omega'_{i} \\
\cross \bra{\omega_{s}-\omega'_{s},\omega_{i}-\omega'_{i}}\hat{\rho}\ket{\omega_{s}+\omega'_{s},\omega_{i}+\omega'_{i}}.
 \end{multline}
 Their marginals correspond to the Joint spectral intensity (JSI), the joint temporal intensity (JTI) or cross-marginal (JTSI):
\begin{align}
 \iint  d\omega_{s} d\omega_{i} W_{\hat{\rho}}(\omega_{s},\omega_{i},t_{s},t_{i})= \text{JTI}(t_{s},t_{i}), \nonumber \\
  \iint dt_{s} dt_{i} W_{\hat{\rho}}(\omega_{s},\omega_{i},t_{s},t_{i})= \text{JSI}(\omega_{s},\omega_{i}),\nonumber \\
    \iint d\omega_{s} dt_{i} W_{\hat{\rho}}(\omega_{s},\omega_{i},t_{s},t_{i})= \text{JTSI}(t_{s},\omega_{i}) \nonumber \\
        \iint d\omega_{i} dt_{s} W_{\hat{\rho}}(\omega_{s},\omega_{i},t_{s},t_{i})= \text{JTSI}(t_{i},\omega_{s}). \nonumber
\end{align}

These four marginals were measured in \cite{PhysRevA.100.033834}. Also in the two-photon case, the issue of Gaussian and non-Gaussian states appear. It was shown in \cite{douce_direct_2013} that the biphoton Wigner function associated to variable $\omega_-$ can be directly measured using the Hong-Ou-Mandel experiment, and also that non-Gaussianity - or, equivalently, a coincidence detection probability greater than $1/2$ - is a proof of frequency entanglement between in the produced photon pair \cite{eckstein_broadband_2008}. 

A final issue to be discussed concerns the efficiency of production and detection of single photons. We can mention deterministic and bright single photon sources based on quantum dots \cite{nowak_deterministic_2014,he_deterministic_2017,somaschi_near-optimal_2016}  that can be integrated on chip  \cite{uppu_-chip_2020} or probabilistic ones, as the one produced by SPDC  \cite{belhassen_-chip_2018,guo_parametric_2017} or by four-wave mixing \cite{paesani_near-ideal_2020}. The state-of-the-art single photon detectors in the 1550 nm range, for instance, can reach 90 \% for superconducting nanowire single-photon detector (SNSPD) composed of materials which can be integrated on chip \cite{you_superconducting_2020}. While the SNSPD used for detecting 925 nm single photons produced by micro-pillar sources \cite{somaschi_near-optimal_2016} reach an efficiency of 60-82 \% in the boson sampling experiment \cite{wang_boson_2019}.

 Also, the CV measurement and manipulation process, which is necessary, for instance, in  teleportation or in the measurement-based protocol, leads to deformation of the spectral or temporal shape of the single photon. This can be associated to encoding errors that reduce  the fidelity of the CV operations. Nevertheless, this is expected, and intrinsic to any type of  CV encoding \cite{pirandola_quantum_2006}.

\section{Discussion and conclusion}

We have described in detail how to encode quantum information in time and frequency of single photons, formally treated as quantum continuous  variables. For this, we built an analogy between these variables and the position and momentum of a particle or the field's quadratures. This analogy was possible because, in the single photon regime, time and frequency operators obey the canonical commutation relations of the Heisenberg algebra, and can be formally manipulated as position and momentum are. As a consequence, one can define a finite set of universal gates that when combined lead to unitary operations that are arbitrary polynomials in time and frequency variables. 

The set of universal operations defined using time and frequency consists of highly non-linear transformations at the single photon level. While the single photon operations in the circuit model can be implemented with current technology, the two-photon interaction is, for the time being, easier to implement using the natural energy and momentum conservation relations observed in some currently used sources of photon pairs. One can also benefit from a complete measurement-based strategy for devices where pump engineering is possible.

We also briefly discussed the equivalent to the dynamics of the single photons evolving under the analogous of a Hamiltonian defined in frequency and time variables. It corresponds to the propagation in a dispersive linear device. Finally, we showed that it is possible to represent and interpret non-classical and symmetry aspects of quantum states using phase space, even though some properties as non-Gaussianity play a less distinctive role than in quadrature-based set-ups in what concerns their physical implementation. 

Our results are relevant from a fundamental point of view, pointing out an original way to interpret, manipulate and use frequency and time state of single photons. They may help developing novel ways to encode quantum information and implement quantum communication and quantum metrology protocols in practice. 

\bibliography{biblioUQC}

\begin{thebibliography}{65}
\bibitem[{Not()}]{Note}
\bibinfo{howpublished}{Note that it is possible to introduce time and frequency
  displacement operator manipulating classical fields obeying the Weyl's
  algebra. This is a consequence of the Fourier transform relation between
  canonically conjugated variable but first, the Planck's constant does not
  intervene in such a case.}
\end{thebibliography}

\appendix

\section{}\label{Expansion}

\subsection{Canonical Commutation Relation}
We now provide the technical details for the calculation of the commutation relations between different unitary operators, polynomials in $\hat \omega$ or $\hat t$. 

To begin with, we compute the commutation relation $\left [ \hat \omega, \hat t \right ]$ using that : 
\begin{eqnarray}\label{omega}
&&\hat \omega = \int \omega \hat a^{\dagger}(\omega)\hat a(\omega){\rm d}\omega \nonumber \\
&&\hat t = \int t \hat a^{\dagger}(t)\hat a(t){\rm d}t,
\end{eqnarray}
with $\hat a(t)=\int \hat a(\omega)e^{-i\omega t}{\rm d}\omega$. For that, we'll use the fact that we are in the single photon regime, which is an intrinsically single mode state. It can thus be expressed in some mode $\hat b_n=\int U_n(\omega)\hat a(\omega){\rm d}\omega$, so that $\hat b_n=\iint U_n(\omega)\hat a(t)e^{i\omega t}{\rm d}\omega {\rm d}t =  \int \tilde U_n(t)\hat a(t){\rm d}t $, where $\tilde U_n(t)=\int U_n(\omega)e^{i\omega t}{\rm d}\omega$ and  $U_n(\omega)$ is an isometric linear mode transformation fulfilling :
$\sum_n U(\omega)^\dagger U_n(\omega') = \delta(\omega-\omega')$. We can then re-write Eq.~(\ref{omega}) in terms of these modes as:
\begin{eqnarray}\label{omegamode}
&&\hat \omega = \sum_{n,m}\int \omega \left ( U_n(\omega)\hat b^{\dagger}_n \right )\left (U^*_m(\omega)\hat b_m \right ){\rm d}\omega \nonumber \\
&&\hat t = \sum_{n,m}\int t \left ( \tilde U_n(t)\hat b^{\dagger}_n \right )\left (\tilde U^*_m(t)\hat b_m \right ){\rm d}t \nonumber,
\end{eqnarray}
so that
\begin{eqnarray}\label{commute}
&&\left [ \hat \omega, \hat t\right ] = \sum_{n,m, n',m'} \iint \omega t U_n(\omega)U^*_m(\omega)\tilde U_{n'}(t)\tilde U^*_{m'}(t)\times \nonumber \\
&&(\hat b^{\dagger}_n\hat b_mb^{\dagger}_{n'}\hat b_{m'}-b^{\dagger}_{n'}\hat b_{m'}\hat b^{\dagger}_n\hat b_m){\rm d}t {\rm d}\omega.
\end{eqnarray}
We can change this sum into
\begin{eqnarray}\label{commute2}
&&\left [ \hat \omega, \hat t\right ] = \nonumber  \\
&&\sum_{n,m, n',m'} \iint \omega t \left(U_n(\omega)U^*_m(\omega)\tilde U_{n'}(t)\tilde U^*_{m'}(t)-\right. \nonumber \\
&&\left. U_{n'}(\omega)U^*_{m'}(\omega)\tilde U_{n}(t)\tilde U^*_{m}(t)\right)
\hat b^{\dagger}_n\hat b_mb^{\dagger}_{n'}\hat b_{m'}{\rm d}t {\rm d}\omega.
\end{eqnarray}
Since $\hat b^{\dagger}_n\hat b_mb^{\dagger}_{n'}\hat b_{m'} = \hat b^{\dagger}_n \hat b_{m'} \delta_{m,n'}+\hat b^{\dagger}_nb^{\dagger}_{n'}\hat b_m\hat b_{m'}$, Eq.~(\ref{commute2}) becomes
\begin{eqnarray}\label{commute3}
&&\left [ \hat \omega, \hat t\right ] = \\
&&\sum_{n,m, m'} \iint  \omega t \left(U_n(\omega)U^*_m(\omega)\tilde U_{m}(t)\tilde U^*_{m'}(t)-\right. \nonumber \\
&&\left. U_{m}(\omega)U^*_{m'}(\omega)\tilde U_{n}(t)\tilde U^*_{m}(t)\right)\hat b^{\dagger}_n\hat b_{m'}{\rm d}t {\rm d}\omega + \nonumber \\
&&\sum_{n,m, n',m'} \iint \omega t \left(U_n(\omega)U^*_m(\omega)\tilde U_{n'}(t)\tilde U^*_{m'}(t)-\right.\nonumber \\
&&\left.U_{n'}(\omega)U^*_{m'}(\omega)\tilde U_{n}(t)\tilde U^*_{m}(t)\right)\hat b^{\dagger}_nb^{\dagger}_{n'}\hat b_m\hat b_{m'}{\rm d}t {\rm d}\omega.
\end{eqnarray}

In the single mode situation, we have that $n=m'=m=n'=k$, so it's clear that the second sum in Eq.~(\ref{commute3}) is equal to zero. We will then focus on the first one. After writing  the $\tilde{U}_m(t)$ as a function of $U_m(\omega')$, the integral over $\omega'$ in the first term of the sum can be integrated by parts. In addition, we use that $\int t e^{-i(\omega'-\omega'')t}{\rm d}t=i\frac{\rm d}{\rm d \omega' }\delta(\omega'-\omega'')$. It leads to 
\begin{equation}\label{firstterm1}
i \sum_n \int  \int \omega  U_k(\omega)(\frac{\rm d}{\rm d \omega' }U_k^*(\omega')) U_n^*(\omega)U_n(\omega') {\rm d}\omega {\rm d}\omega'.
\end{equation}
We now use the orthogonality condition:  $\sum_n  U_n^*(\omega)U_n(\omega') = \delta(\omega-\omega')$, leading to 
\begin{equation}\label{firstterm2}
i  \int \omega  U_k(\omega)(\frac{\rm d}{\rm d \omega }U_k^*(\omega))  {\rm d}\omega,
\end{equation}

which can again be integrated by parts to obtain
\begin{equation}\label{firstterm3}
i  + i\int \omega  U_k^*(\omega)(\frac{\rm d}{\rm d \omega }U_k(\omega))  {\rm d}\omega. 
\end{equation}
We now analyse the second part of the first term of Eq.~(\ref{commute3}) and proceed to an integration by parts as in Eq.~(\ref{firstterm1}), which leads to
\begin{equation}\label{secondterm1}
i \sum_n \int  \int \omega  U_k^*(\omega)(\frac{\rm d}{\rm d \omega' }U_k(\omega')) U_n(\omega)U_n^*(\omega') {\rm d}\omega {\rm d}\omega'.
\end{equation}
We can again perform the sum in $n$ to reach
\begin{equation}\label{secondterm1}
i   \int \omega  U^*_k(\omega)(\frac{\rm d}{\rm d \omega }U_k(\omega)) {\rm d}\omega,
\end{equation}
that will cancel the last term in Eq.~(\ref{firstterm3}). We have then
\begin{equation}\label{comutador}
\left [ \hat \omega, \hat t \right ] = i \hat b^{\dagger}_k \hat b_k = i\mathds{I}, 
\end{equation}
since $\hat b^{\dagger}_k \hat b_k$  is the photon number operator in the single photon mode denoted here as $k$. This operator is equal to the identity in the single photon subspace, since there is only one single photon in each mode. Notice that no assumption on the state was made. These are general properties of one photon per mode states.

\subsection{Unitary operators}

A given unitary operator which is generated by a hermitian operator which is a polynomial $\hat \omega$ (or, equivalently, $\hat t$), can be expressed in the single photon subspace ${\cal S}_1$ as $e^{i\hat \omega_{a}^n \alpha}$. Since 
\begin{equation}\label{Exp}
e^{i\hat \omega_{a}^n \alpha}= \sum_k^{\infty} \frac{(i\alpha)^k}{k!} \hat \omega^{nk},
\end{equation}
we will focus on the term $\hat \omega^{nk} \equiv \hat \omega^s$. Since $\hat \omega^s= \hat \omega^2 \hat \omega^{(s-2)}$, we have that : 
\begin{equation}\label{expansion}
\hat \omega_{a}^2=\left ( \int \omega \hat a^{\dagger}(\omega)\hat a(\omega){\rm d}\omega \right )^2.
\end{equation}
The r.h.s. of Eq.~(\ref{expansion}) can be easily computed, leading to $\iint \omega \omega' \hat a^{\dagger}(\omega)\hat a(\omega)\hat a^{\dagger}(\omega')\hat a(\omega'){\rm d}\omega {\rm d}\omega'$. Using that $\hat a(\omega)\hat a^{\dagger}(\omega')=\hat a^{\dagger}(\omega')\hat a(\omega) + \delta(\omega-\omega')$, we obtain 
\begin{eqnarray}\label{expansion2}
&&\hat \omega_{a}^2= \iint  \omega \omega' \hat a^{\dagger}(\omega) \hat a^{\dagger}(\omega')\hat a(\omega)\hat a(\omega'){\rm d}\omega {\rm d}\omega' \nonumber \\
&&+ \int \omega^2 \hat a^{\dagger}(\omega)\hat a(\omega){\rm d}\omega. 
\end{eqnarray}
However, the first term in the r.h.s. of Eq.~(\ref{expansion2}) is equal to zero in the single photon subspace. We have thus :

\begin{equation}\label{n}
\hat \omega_{a}^2= \int \omega^2 \hat a^{\dagger}(\omega)\hat a(\omega){\rm d}\omega. 
\end{equation}

It's thus clear that $\hat \omega_{a}^s= \int \omega^s \hat a^{\dagger}(\omega)\hat a(\omega){\rm d}\omega$ and, consequently, 

\begin{equation}\label{ExpFin}
e^{i\hat \omega_{a}^n \alpha}=  \int e^{i \omega^n\alpha} \hat a^{\dagger}(\omega)\hat a(\omega){\rm d}\omega.
\end{equation}

Using Eq.~(\ref{ExpFin}) and the commutation relation between creation and annihilation operators, we can prove all the commutation relations appearing in this manuscript. Of course, the same reasoning is valid for unitary operators involving the time operator. 

The particular case of $n=1$ is relevant for the definition of the displacement operators. In this case, Eq.~(\ref{ExpFin}) becomes : 
\begin{eqnarray}\label{Disp}
&&e^{i\hat \omega_{a} \alpha}=  \int e^{i \omega \alpha} \hat a^{\dagger}(\omega)\hat a(\omega){\rm d}\omega =\\
&&   \iiint  e^{i \omega \alpha} \hat a^{\dagger}(t)e^{-i \omega t}\hat a(t')e^{i \omega t'}{\rm d}\omega {\rm d}t {\rm d}t',\nonumber
\end{eqnarray}
which, after integration in $\omega$ leads to : 
\begin{equation}\label{Disp2}
e^{i\hat \omega_{a} \alpha}=  \int  \hat a^{\dagger}(t+\alpha)\hat a(t){\rm d}t.
\end{equation}

\end{document}